# Generalized Sigma Model Description of the Light $J = 0$ Mesons [*]

Amir H. Fariborz [a] [‡], Renata Jora [b] [†], and Joseph Schechter [b] [§]

[a] *Department of Mathematics/Science, State University of New York Institute of Technology, Utica, NY 13504-3050, USA*
[b] *Department of Physics, Syracuse University, Syracuse, NY 13244-1130, USA*

Within a linear sigma model framework, possible mixing between two chiral nonets (a two quark nonet, and a four quark nonet) below 2 GeV is studied. Incorporating the U(1)$_A$ behavior of the underlying QCD, and working in the isospin invariant limit, the mass spectra of the $I = 0$, $I = 1/2$ and $I = 1$ pseudoscalars, and the $I = 1/2$ scalars are studied, and estimates of their quark content are presented. It is found, as expected, that the ordinary and the excited pseudoscalars generally have much less two and four quark admixtures compared to the respective scalars. As by-products, several quantities such as the four quark vacuum condensate, and the decay constant of excited states are predicted.

## I. INTRODUCTION

To understand the low-energy QCD, we need to explore the properties of the light hadrons, and that means we should be able to answer questions such as:

- What are the light hadrons made of? Do they follow the simplest quark structures (i.e. $\bar{q}q$ for mesons and $qqq$ for baryons), or they are more complex, and may have, for example, exotic structures (such as four quark mesons and pentaquark baryons). Or these states may even be more complex and have a hybrid structure.

- What are their measurable properties such as their mass and decay width? For example, some states (like scalars $f_0(600)$ or $\kappa(900)$) have not been well established experimentally, and therefore, theoretical work is needed to explore their properties. Of course, the properties of these states are closely related to their internal substructure, and therefore, these two questions are practically two sides of the same coin.

Among mesons, light scalars have been particularly difficult to understand [1]. States such as a light and broad $I = J = 0$ sigma resonance in the 500-600 MeV region [2], as well as a light and broad $I, J = 1/2, 0$ kappa resonance in the 700-900 MeV region [3] have been at the focus of many theoretical and experimental investigation. Together with the well established $f_0(980)$ and $a_0(980)$ scalar resonances, these comprise a putative nonet of "elementary particles." Furthermore, this nonet seems likely to have a quark structure like $qq\bar{q}\bar{q}$ rather than the conventional $q\bar{q}$ [4, 5]. This of course raises the question of where are the conventional $q\bar{q}$ p-wave scalars expected in the quark model. Arguments have been given [6–10] that the experimental data are better fit when the two scalar nonets mix with each other and the resulting "level repulsion," pushes the conventional scalars to higher masses than otherwise expected.

In order to further explore the feature of mixing between $q\bar{q}$ type and $qq\bar{q}\bar{q}$ type states it seems interesting to consider a linear SU(3)× SU(3) sigma model which contains also the pseudoscalar nonet partners of these two scalar nonets. The SU(2) linear sigma model was first given in ref. [11]. It was used as a basis for understanding the current algebra treatment of $\pi\pi$ scattering near threshold in ref. [12]. The SU(3) version was given in the first of ref. [13]. A detailed application to the low energy pseudoscalar mass spectrum was given [14] before QCD in which, among other things, it was shown how a U(1)$_A$ violating term natural in the SU(3) model could solve the $\eta'$ problem. Such a term was later discovered to arise from instanton effects [15]. The connection was pointed out in ref. [16] and emphasized by 't Hooft [17].

The model containing two different chiral nonets to be discussed here was first proposed in section V of ref. [7] and an initial treatment, neglecting flavor symmetry breaking, was given. See also [18] for a related study. Here we present the formalism for treating consequences of the model which hold (at tree level) just due to the symmetry structure of the model and will give a numerical treatment using what might be the simplest choice of symmetry breaking terms. More details of the model can be found in a previous work [19]. Some new results are given in the present work.

---

## II. BRIEF DESCRIPTION OF THE MODEL

We consider two chiral nonets of mesons: A two-quark chiral nonet $M$ and a four-quark chiral nonet $M'$. [22] At the level of effective chiral Lagrangian, $M$ and $M'$ are distinguished from each other by their different U(1)$_A$ transformation properties. The structure of these nonets in terms of the underlying quark fields, and the subsequent chiral transformation properties, have been discussed in detail [19]. These fields may be decomposed into Hermitian scalar (S) and pseudoscalar ($\phi$) nonets as,

$$M = S + i\phi \quad , \quad M' = S' + i\phi'. \tag{1}$$

We take the vacuum values of the diagonal components of $S$ and $S'$ as

$$\langle S_a^b \rangle = \alpha_a \delta_a^b, \qquad \langle S_a'^b \rangle = \beta_a \delta_a^b. \tag{2}$$

In the iso-spin invariant limit, $\alpha_1 = \alpha_2$ and $\beta_1 = \beta_2$ while in the SU(3) invariant limit, $\alpha_1 = \alpha_2 = \alpha_3$ and $\beta_1 = \beta_2 = \beta_3$.

The Lagrangian density which defines our model is

$$\begin{aligned}\mathcal{L} = & -\frac{1}{2}\mathrm{Tr}\left(\partial_\mu M \partial_\mu M^\dagger\right) - \frac{1}{2}\mathrm{Tr}\left(\partial_\mu M' \partial_\mu M'^\dagger\right) - V_0\left(M, M'\right) + \mathrm{Tr}\left[A(M + M^\dagger)\right] \\ & + \frac{1}{288c}[\ln(\frac{\det M}{\det M^\dagger})]^2, \end{aligned} \tag{3}$$

where $V_0(M, M')$ stands for a general function made from SU(3)$_L \times$ SU(3)$_R$ (but not necessarily U(1)$_A$) invariants formed out of $M$ and $M'$. The fourth term breaks the flavor symmetry and is chosen to mock up the quark mass terms, with $A = \mathrm{diag}(A_1, A_2, A_3)$. The last terms mocks up the anomalous U(1)$_A$ equation of QCD.

The underlying chiral symmetry leads to a set of generating equations from which a number of relationships among masses are obtained [19]. This gives the matrix of squared $\pi$ and $\pi'$ masses

$$(M_\pi^2) = \begin{bmatrix} x_\pi + z_\pi^2 y_\pi & -z_\pi y_\pi \\ -z_\pi y_\pi & y_\pi \end{bmatrix}. \tag{4}$$

where $x_\pi = 2A_1/\alpha_1$, $y_\pi = \langle \partial^2 V / \partial \phi'^1_2 \partial \phi'^2_1 \rangle$, and $z_\pi = \beta_1/\alpha_1$. It is clear that $z_\pi$ is a measure of the mixing between $\pi$ and $\pi'$ since the matrix becomes diagonal in the limit when $z_\pi$ is set to zero. So we see that $x_\pi$ would be the squared pion mass in the single M model and $y_\pi$ represents the squared mass of the "bare" $\pi'$. In this system, there are four unknowns $\alpha_1$, $\beta_1$, $A_1$ and $y_\pi$. The experimental inputs are $m_\pi$ and $m_{\pi'}$, together with $F_\pi$, which is calculated in terms of these unknowns: $F_\pi = (\alpha_1 + \alpha_2)\cos\theta_\pi - (\beta_1 + \beta_2)\sin\theta_\pi$. The transformation between the diagonal fields (say $\pi^+$ and $\pi'^+$) and the original pion fields is defined as:

$$\begin{bmatrix} \pi^+ \\ \pi'^+ \end{bmatrix} = \begin{bmatrix} \cos\theta_\pi & -\sin\theta_\pi \\ \sin\theta_\pi & \cos\theta_\pi \end{bmatrix} \begin{bmatrix} \phi_1^2 \\ \phi'^2_1 \end{bmatrix}. \tag{5}$$

Therefore, for the pion system, we run one of the parameters (we take it to be $x_\pi$) and all other unknowns are determined. The same situation holds for the kaon system: The mass matrix has the same structure as (4) with four unknowns $A_3$, $\alpha_3$, $\beta_3$ and $\langle \partial^2 V / \partial(\phi'^1_3 \partial \phi'^3_1)\rangle$, and there are three experimental inputs $m_K$, $m_{K'}$, and $F_K$. Therefore, the parameters in the pion and kaon systems are all determined in terms of two parameters ($x_\pi$ and $x_k = 2(A_3 + A_1)/(\alpha_3 + \alpha_1)$). It is shown [19] that when these parameters are fed into the kappa system, the chiral symmetry imposes a condition between $x_\pi$ and $x_k$. Together with the experimental inputs for $m_\kappa$ and $m_{\kappa'}$, the pion, kaon and kappa systems are predicted in terms of only one unknown that we take to be $x_\pi$. To judge which $x_\pi$ is favored, we look into the $\eta$ system which is more complicated because, even in the isotopic spin invariant limit, there are four different $I = 0$ pseudoscalars which can mix with each other. The part of the Lagrangian describing the masses of the $I = 0$ pseudoscalars is then: $\mathcal{L} = -(1/2)\Phi_0^T(M_\eta^2)\Phi_0$, where $(M_\eta^2)$ is a symmetric $4 \times 4$ matrix and:

$$\Phi_0^T = \begin{bmatrix} (\phi_1^1 + \phi_2^2)/\sqrt{2} & \phi_3^3 & (\phi'^1_1 + \phi'^2_2)/\sqrt{2} & \phi'^3_3 \end{bmatrix}. \tag{6}$$

This will introduce four additional parameters that we take to be: $c$ in Lagrangian (3), $\partial^2 V / (\partial \phi'^1_1 \partial \phi'^1_1)$, $\partial^2 V / (\partial \phi'^1_1 \partial \phi'^3_3)$, and $\partial^2 V / (\partial \phi'^3_3 \partial \phi'^3_3)$.

Next we incorporate the experimental data to make predictions about these four systems (pions, kaons, kappas, and etas).

## III. RESULTS AND DISCUSSION

We take the well known lowest pseudoscalar nonet masses and decay constants (in GeV): $m_\pi = 0.137$, $m_K = 0.496$, $m_\eta = 0.548$, $m_{\eta'} = 0.958$, $F_\pi = 0.131$, and $F_K = 0.160$. For the excited states, we consider the next-to-lowest states listed in PDG [20]: $m_{\pi'} = 1.30$, $m_{K'} = 1.46$, $m_\kappa = 0.90$, and $m_{\kappa'} = 1.42$. For the excited $\eta$ type states below 2 GeV, the possible masses (all in GeV) are: 1.294, 1.410, 1.476, 1.760. It seems more difficult to a priori choose which are most relevant so we shall study all possible pairings in a systematic way.

First we consider the $\pi - \pi'$, $K - K'$ and $\kappa - \kappa'$ systems. As already discussed, by specifying $x_\pi$, all features of these systems in our model will be determined. Table I shows the predicted physical parameters for three values of $x_\pi$. The "quark mass ratio", $A_3/A_1 = 30.3$ in the single $M$ model is not very different from the value of 31.3 obtained using the values in the $x_\pi = 0.019 \text{ GeV}^2$ column. The $\bar{q}q$ meson condensates $\alpha_1$ and $\alpha_3$ are also very similar. Of course the "four quark" meson condensates $\beta_1$ and $\beta_2$ are zero without $M'$. Despite the similarities, the 6.4° mixing angle already corresponds to about ten percent "four quark" admixture in the physical pion wave function. Considering that the accuracy of current algebra predictions for low energy pion physics is roughly ten percent, it seems that this choice of $x_\pi$ is the most plausible one. One sees from the second and third columns that relatively small increases in $x_\pi$ lead to large increases in four quark admixture for the pion and the kaon. Interestingly, the behavior of the four quark admixture in the strange scalar meson $\kappa$ is very different. When the pseudoscalars are closer to pure "two quark" states in the model the scalar has a large four quark admixture (32.4°, with the choice of $x_\pi$ in the first column). Thus the result is consistent with having a fairly large four quark component in the light scalars. The analogs of the two quark condensates $\alpha_1 = \alpha_2$ and $\alpha_3$ are approximately equal, in agreement with the usual assumption that the vacuum is approximately SU(3) symmetric. The analogs of the four quark condensates in this model are roughly an order of magnitude smaller than the similarly normalized two quark condensates. They are furthermore seen to deviate appreciably from SU(3) symmetry.

To check the consistency of the model and our initial assumption of $M$ being a two-quark chiral nonet and $M'$ being a four-quark chiral nonet, we use the result of the fits and extract the "bare" masses for the pions, kaons and kappas. This is given in Table II. We see that the bare masses of the pseudoscalars in $M$ are less than the corresponding masses in $M'$. However, the situation is reversed for the scalar kappas, which is understandable based on the hyperfine interaction for four-quark scalars discovered by Jaffe [4].

|  | $x_\pi$=0.019 | $x_\pi$=0.021 | $x_\pi$=0.022 |
|---|---|---|---|
| $\theta_\pi$ (deg.) | $-6.37$ | $-19.1$ | $-22.7$ |
| $\theta_k$ (deg.) | $-11.2$ | $-22.9$ | $-26.2$ |
| $\theta_\kappa$ (deg.) | 34.1 | 28.1 | 26.5 |
| $A_1(\text{GeV}^3)$ | $6.19 \times 10^{-4}$ | $6.51 \times 10^{-4}$ | $6.66 \times 10^{-4}$ |
| $A_3(\text{GeV}^3)$ | $1.94 \times 10^{-2}$ | $2.07 \times 10^{-2}$ | $2.12 \times 10^{-2}$ |
| $\alpha_1$ (GeV) | $6.51 \times 10^{-2}$ | $6.20 \times 10^{-2}$ | $6.06 \times 10^{-2}$ |
| $\alpha_3$ (GeV) | $9.24 \times 10^{-2}$ | $8.83 \times 10^{-2}$ | $8.69 \times 10^{-2}$ |
| $\beta_1$ (GeV) | $7.18 \times 10^{-3}$ | $2.12 \times 10^{-2}$ | $2.50 \times 10^{-2}$ |
| $\beta_3$ (GeV) | $2.03 \times 10^{-2}$ | $3.38 \times 10^{-2}$ | $3.74 \times 10^{-2}$ |

TABLE I: Mixing angles, vacuum condensates and flavor symmetry breaking parameters, for pion, kaon and kappa systems, found by a fit of the model to experimental masses and decay widths, for three values of $x_\pi$ (GeV$^2$).

Now consider the mixing of the four $\eta$ type fields in this model. As discussed in the previous section, there are, after using the symmetry information, four new unknown parameters characterizing the $\eta$ system. Thus taking the four mass eigenvalues from experiment could in principle determine, together with results from the $\pi - \pi'$, $k - k'$ and $\kappa - \kappa'$ systems, everything about the $\eta$ system for a given value of $x_\pi$. However there is no guarantee that there will be an exact solution for all choices of experimental parameters. This is the case, in fact, so we will search numerically for a choice of "theoretical" masses which will best fit the experimental inputs. The criterion for goodness of fit will be taken to be the smallness of the quantity:

$$\chi \equiv \sum_i |m_i^{\text{exp.}} - m_i^{\text{theo.}}| / m_i^{\text{exp.}}. \tag{7}$$

There are three established candidates and one not yet established candidate below 2 GeV for the two excited $\eta$ states. This yields six possible scenarios for choosing them. The quantity $\chi$ for each choice is shown in Table III for three values of the parameter $x_\pi$. It may be observed that the fits typically get worse with increasing $x_\pi$, so it is reasonable to consider the choice 0.019 GeV$^2$ for this quantity as we did previously. Figure 1 shows the two and four quark components of the four physical etas. In scenarios 3, 5 and 6 [involving the $\eta(1760)$], the $\eta(958)$ unexpectedly acquires a large four-quark component, and seems inconsistent (see Fig.1). On the other hand, scenarios 1, 2 and 4 seem more consistent: $\eta(548)$ has a large $s\bar{s}$ component; the $\eta(958)$ and the $\eta_3$ [which is either $\eta(1295)$ or $\eta(1405)$] have comparable two and four quark components; and finally the $\eta_4$ [which is either $\eta(1405)$ or $\eta(1475)$] is mainly a four quark state with a negligible two quark admixtures. The large four quark component of $\eta(958)$ may be an artifact of not having a glue component [which is known to be large for $\eta(958)$] in this analysis and will be investigated in future works. The smallest value of $\chi$ is for scenario 2. Even though the values of $x_\pi = 0.021$ GeV$^2$ and $x_\pi = 0.022$ GeV$^2$ are not favored, but for completeness and comparison, the six scenarios of Table III for these two values of $x_\pi$ are presented in Figs. 2 and 3. We can immediately observe that these figures too illustrate that large mixings are disfavored. For example, we can see that with such large mixings, the $\eta(958)$ has a dominant four quark components which is not supported by the conventional phenomenology.

|   | $x_\pi$=0.019 | $x_\pi$=0.021 | $x_\pi$=0.022 |
|---|---|---|---|
| $m_\pi^0$ | 0.198 | 0.446 | 0.517 |
| $m_{\pi'}^0$ | 1.29 | 1.23 | 1.20 |
| $m_k^0$ | 0.563 | 0.729 | 0.783 |
| $m_{k'}^0$ | 1.44 | 1.36 | 1.33 |
| $m_\kappa^0$ | 1.29 | 1.33 | 1.34 |
| $m_{\kappa'}^0$ | 1.09 | 1.04 | 1.03 |

TABLE II: "Brae" masses (in GeV) of pions, kaons and kappas for the three values of $x_\pi$ (in GeV$^2$). (The "bare" masses refer to the unmixed masses, which are the diagonal elements of the mass matrix.)

| Scenario | $x_\pi$=0.019 | $x_\pi$=0.021 | $x_\pi$=0.022 |
|---|---|---|---|
| 1:$\{\eta(1295), \eta(1405)\}$ | $6.23 \times 10^{-2}$ | $3.99 \times 10^{-1}$ | $5.08 \times 10^{-1}$ |
| 2:$\{\eta(1295), \eta(1475)\}$ | $2.85 \times 10^{-2}$ | $3.39 \times 10^{-1}$ | $4.44 \times 10^{-1}$ |
| 3:$\{\eta(1295), \eta(1760)\}$ | $2.35 \times 10^{-2}$ | $1.37 \times 10^{-1}$ | $2.28 \times 10^{-1}$ |
| 4:$\{\eta(1405), \eta(1475)\}$ | $8.28 \times 10^{-2}$ | $3.63 \times 10^{-1}$ | $4.49 \times 10^{-1}$ |
| 5:$\{\eta(1405), \eta(1760)\}$ | $1.50 \times 10^{-2}$ | $1.62 \times 10^{-1}$ | $2.38 \times 10^{-1}$ |
| 6:$\{\eta(1475), \eta(1760)\}$ | $2.84 \times 10^{-2}$ | $1.78 \times 10^{-1}$ | $2.68 \times 10^{-1}$ |

TABLE III: The goodness of fit for 6 possible scenarios and for three values of $x_\pi$ (GeV$^2$). Each scenario corresponds to a choice of $\eta$ type fields including the $\eta(548)$ and the $\eta(958)$ as well as the two listed in the left hand column.

To sum up, the value $x_\pi = 0.019$ GeV$^2$ leads to fairly small four quark content in the light pseudoscalars $\pi, K, \eta$, at the same time the light scalar $\kappa$ has an appreciable four quark component. The "excited" $\eta$'s are predominantly four quark states. The $\eta(958)$ is mainly two quark in content but has a non trivial four quark piece. The results obtained here provide supporting evidence for the feature, illustrated in the first treatment of this model [7], that the lightest scalars, unlike the lightest pseudoscalars, have appreciable four quark components.

An interesting feature of our model is the presence of "four quark" condensates as signaled by the non-zero values of the $\beta_a$. The estimate is [19]:

$$|\langle ds\bar{d}\bar{s}\rangle| \sim \Lambda_{QCD}^5 \beta_1 \approx 4 \times 10^{-5} \, \text{GeV}^6. \qquad (8)$$

Also, the prediction of our model for the decay constant of excited pion and kaon states, as well as for the kappa and its excited state is given in Table IV. The predicted value of $F_{\pi'}$ around 0.16 MeV is qualitatively consistent with prior theoretical estimate value for the excited pion [21].

We are currently investigating several interesting and closely related issues in this model, including the isospin violation and the mass spectrum of the $I = 1$ scalars, as well as inclusion of the glueballs and their contribution to the mass spectrum of the $I = 0$ scalars and pseudoscalars.

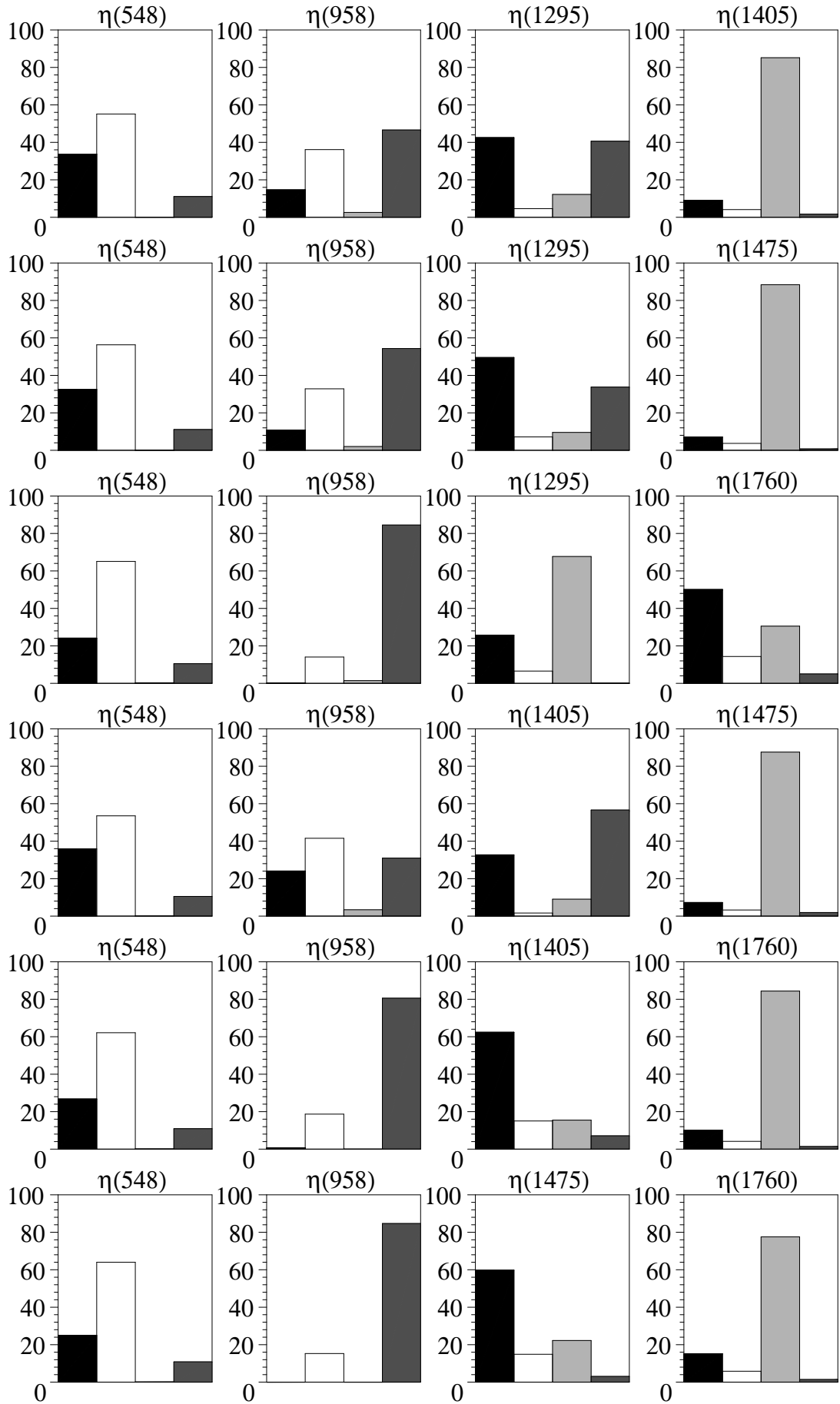

FIG. 1: Graphs in rows one to six respectively represent the scenarios one to six given in Table 1, when $x_\pi = 0.019$ GeV$^2$. The height of the columns give the percentage of the quark component: Black, white, light gray, and dark gray, respectively give the percentage of $(u\bar{u} + d\bar{d})/\sqrt{2}$, $s\bar{s}$, $(\bar{d}s ds + \bar{s}u su)/\sqrt{2}$, and $\bar{u}\bar{d}ud$.

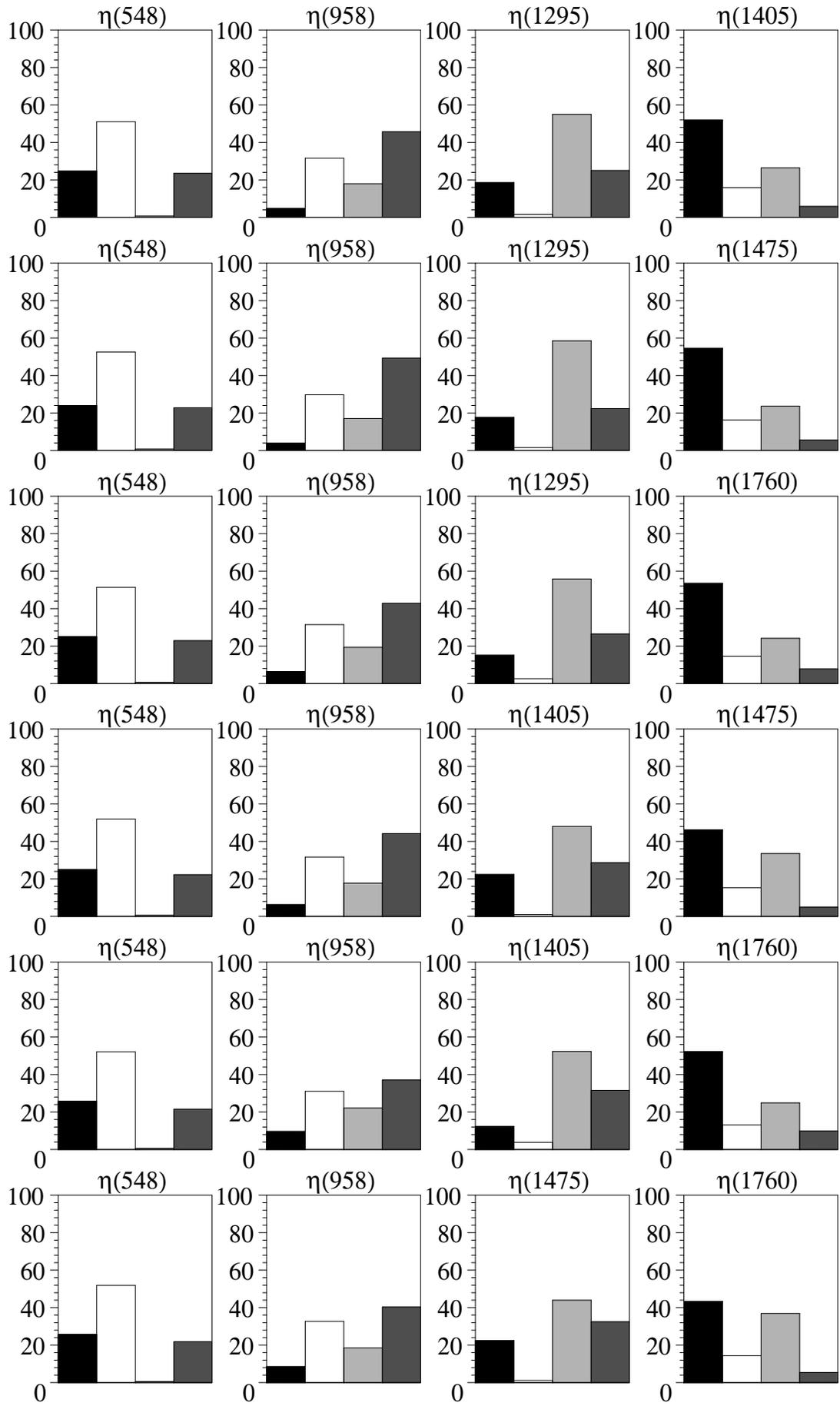

FIG. 2: Graphs in rows one to six respectively represent the scenarios one to six given in Table 1, when $x_\pi = 0.021$ GeV$^2$. The height of the columns give the percentage of the quark component: Black, white, light gray, and dark gray, respectively give the percentage of $(u\bar{u} + d\bar{d})/\sqrt{2}$, $s\bar{s}$, $(\bar{d}\bar{s}ds + \bar{s}\bar{u}su)/\sqrt{2}$, and $\bar{u}\bar{d}ud$.

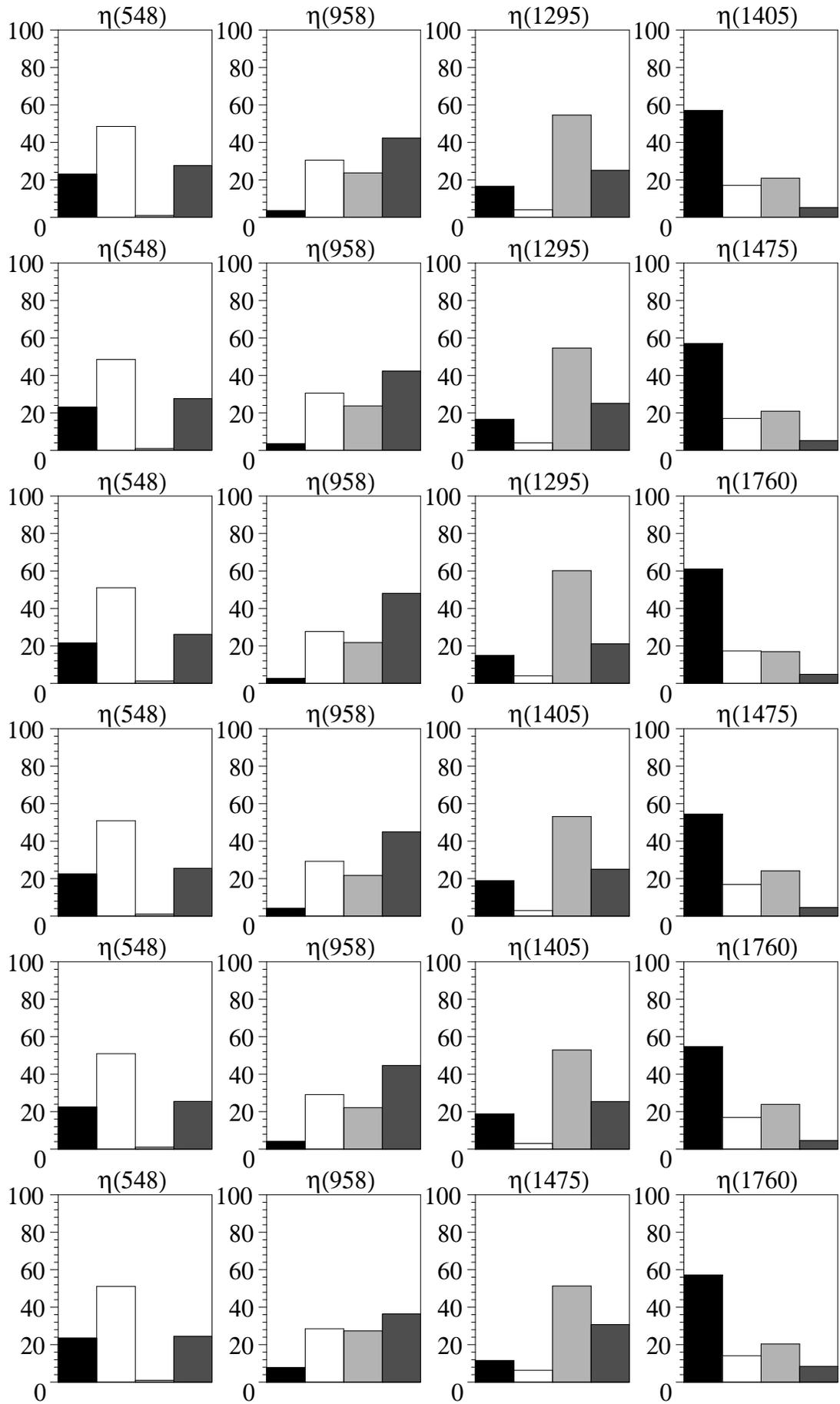

FIG. 3: Graphs in rows one to six respectively represent the scenarios one to six given in Table 1, when $x_\pi = 0.022$ GeV$^2$. The height of the columns give the percentage of the quark component: Black, white, light gray, and dark gray, respectively give the percentage of $(u\bar{u} + d\bar{d})/\sqrt{2}$, $s\bar{s}$, $(\bar{d}s ds + \bar{s}u su)/\sqrt{2}$, and $\bar{u}\bar{d}ud$.

|  | $x_\pi=0.019$ | $x_\pi=0.021$ | $x_\pi=0.022$ |
|---|---|---|---|
| $F_\pi$ | 131 | 131 | 131 |
| $F_{\pi'}$ | −0.162 | −0.505 | −0.608 |
| $F_k$ | 160 | 160 | 160 |
| $F_{k'}$ | −3.65 | −7.80 | −9.06 |
| $F_\kappa$ | 15.3 | 17.3 | 18.0 |
| $F_{\kappa'}$ | 26.1 | 23.5 | 22.8 |

TABLE IV: Predicted decay constants (in MeV) for the three values of $x_\pi$ (in GeV$^2$)

## Acknowledgments

The work of A.H.F. has been supported by a 2005 Crouse Grant from the School of Arts and Sciences, SUNY Institute of Technology. The work of R.J. and J.S. is supported in part by the U. S. DOE under Contract no. DE-FG-02-85ER 40231.